\documentclass[aip,cha,twocolumn,showpacs,preprintnumbers,amsmath,amssymb,superscriptaddress,10pt]{revtex4-1}

\usepackage{colortbl}
\usepackage[table]{xcolor}
\usepackage{tabularx}
\usepackage{amsmath}
\usepackage[latin1]{inputenc}
\usepackage{bm}
\usepackage{graphicx}
\usepackage{SIunits}
\usepackage{subfigure}
\usepackage{enumerate}
\usepackage{multirow}
\usepackage{hyphenat}

\newcommand{\be}{\begin{equation}}
\newcommand{\ee}{\end{equation}}
\newcommand{\ba}{\begin{eqnarray}}
\newcommand{\ea}{\end{eqnarray}}
\newcommand{\ban}{\begin{eqnarray*}}
\newcommand{\ean}{\end{eqnarray*}}

\begin{document}

\title{\LARGE Full-field implementation of a perfect eavesdropper on a quantum cryptography system}

\author{Ilja Gerhardt}
\affiliation{These authors contributed equally to this work}
\affiliation{Centre for Quantum Technologies and Department of Physics, National University of Singapore, 3 Science Drive 2, Singapore 117543}

\author{Qin Liu}
\affiliation{These authors contributed equally to this work}
\affiliation{Department of Electronics and Telecommunications, Norwegian University of Science and Technology, NO-7491 Trondheim, Norway}

\author{Ant{\'i}a Lamas-Linares}
\affiliation{Centre for Quantum Technologies and Department of Physics, National University of Singapore, 3 Science Drive 2, Singapore 117543}

\author{Johannes Skaar}
\affiliation{Department of Electronics and Telecommunications, Norwegian University of Science and Technology, NO-7491 Trondheim, Norway}
\affiliation{University Graduate Center, NO-2027 Kjeller, Norway}

\author{Christian Kurtsiefer}
\email{christian.kurtsiefer@gmail.com}
\affiliation{Centre for Quantum Technologies and Department of Physics, National University of Singapore, 3 Science Drive 2, Singapore 117543}

\author{Vadim~Makarov}
\email{makarov@vad1.com}
\affiliation{Department of Electronics and Telecommunications, Norwegian University of Science and Technology, NO-7491 Trondheim, Norway}

\date{18 March 2012}

\maketitle

{\bf Quantum key distribution (QKD) allows two remote parties to grow a shared secret key. Its security is founded on the principles of quantum mechanics, but in reality it significantly relies on the physical implementation. Technological imperfections of QKD systems have been previously explored, but no attack on an established QKD connection has been realized so far. Here we show the first full-field implementation of a complete attack on a running QKD connection. An installed eavesdropper obtains the entire `secret' key, while none of the parameters monitored by the legitimate parties indicate a security breach. This confirms that non-idealities in physical implementations of QKD can be fully practically exploitable, and must be given increased scrutiny if quantum cryptography is to become highly secure.}

Secret communication provided by cryptography is needed in many activities of the human civilization -- military, commerce, government and private affairs. The long history of cryptography is a continual cat-and-mouse game of cryptographic systems being broken and replaced with new, stronger ones~\cite{singh:99}. Quantum cryptography, as one of the latest techniques, promised for the first time a security which is not based on mathematical conjectures but on the laws of physics~\cite{bennett:84,bennett:92a}. Technologically, quantum cryptography has matured to experiments over $\le$$250$~km distance~\cite{stucki:09}, and several commercial systems are available. Although security of the QKD protocol is unconditionally proven~\cite{gottesman:04,scarani:09}, deviations of actual hardware from the idealized model still present a challenge. Various attacks have been proposed exploiting imperfections of components in QKD scheme: light modulators~\cite{vakhitov:01,xu:10}, photon sources~\cite{felix:01,nauerth:09} and detectors~\cite{kurtsiefer:01a,makarov:06,lamas-linares:07a,zhao:08,makarov:10,lydersen:10,wiechers:10}. However none of these proposals implemented an attack that eavesdropped the secret key, leaving the question of practicality of technological vulnerabilities unresolved.

We picked one of the proposed attack methods, fully implemented the eavesdropper Eve, and used it to attack an installed QKD line. The QKD system under attack is a well-designed one used previously in several experiments~\cite{marcikic:06,ling:08,peloso:09}, and openly documented~\cite{kurtsiefer:08}. We treated QKD hardware and software as `given' and kept all its settings as they had been set for QKD prior to this study. The hardware and software are assumed fully known to Eve, according to Kerckhoffs' principle~\cite{kerckhoffs:1883}.

In this paper, we demonstrate the full-field implementation of this eavesdropping attack in realistic conditions over a 290-m fibre link between the transmitter Alice and the receiver Bob. From multiple QKD sessions over a few hours, Eve obtains the same `secret' key as Bob, while the usual parameters monitored in the QKD exchange are not disturbed, leaving Eve undetected.

\section*{Results}

\subsection{The faked-state attack}

\begin{figure*}
  \includegraphics[width=17cm]{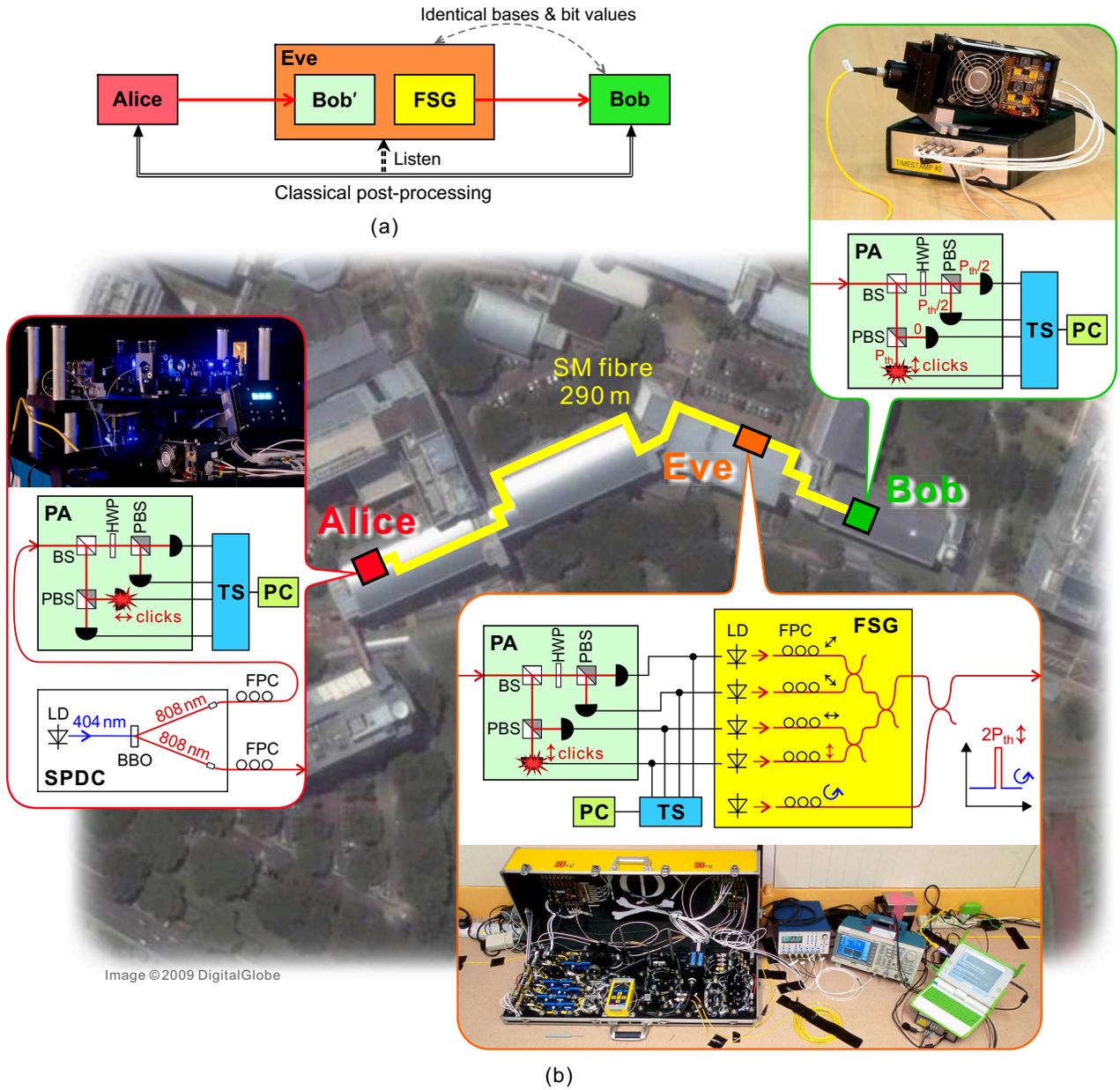}
  \caption{{\bf Eavesdropping experiment}. ({\bf a}) Principle of the faked-state attack. ({\bf b}) Attack on installed QKD system spanning four buildings at the campus of the National University of Singapore. In Alice, polarization-entangled photon pairs were produced in a type-II spontaneous parametric down-conversion (SPDC) source~\cite{marcikic:06,peloso:09}. One photon was measured locally by Alice; the other one was sent through a $290$~m single-mode (SM) fibre line to Bob. Eve was inserted at a mid-way point. All three parties used identical polarization analysers (PA); clicks were registered with timestamp units (TS). Under attack, Bob's detectors clicked controllably when illuminated by an optical pulse with peak power $\geq {\rm P_{th}}$. In the example, to address the target detector for vertically polarized photons, Eve sent a faked state with vertical polarization and peak power ${\rm 2P_{th}}$. Each of Bob's detectors in the conjugate ($45\,^{\circ}$ rotated) basis received a pulse of peak power ${\rm P_{th}/2}$, and thus remained blinded. See Methods Section~\ref{sec:suppl-complete-eve} for a complete description of Eve's setup. In the diagram: BS, 50/50\% beamsplitter; PBS, polarizing beamsplitter; HWP, half-wave plate; FPC, fibre polarization controller; BBO, $\beta$-barium-borate crystal.}
  \label{fig:layout}
\end{figure*}

We have chosen a `faked-state attack' (Fig.~\ref{fig:layout}a)~\cite{makarov:05}. Eve uses a replica of the legitimate receiver unit (Bob$'$) to intercept and measure all quantum states sent by Alice. She further uses a faked-state generator (FSG) to force Bob to output identical bases and bit values, so that Eve and Bob have the same raw key. Eve also records unencrypted communication in the classical channel, and computes the final secret key (identical to Alice's and Bob's) by repeating the same sifting, error correction and privacy amplification procedures~\cite{bennett:92a,scarani:09} as Bob. Unlike the traditional intercept-resend attack~\cite{bennett:84,bennett:92a}, the faked-state attack does not introduce errors in the key and therefore is not detected by the QKD protocol.

\begin{figure}
  \includegraphics[width=8.5cm]{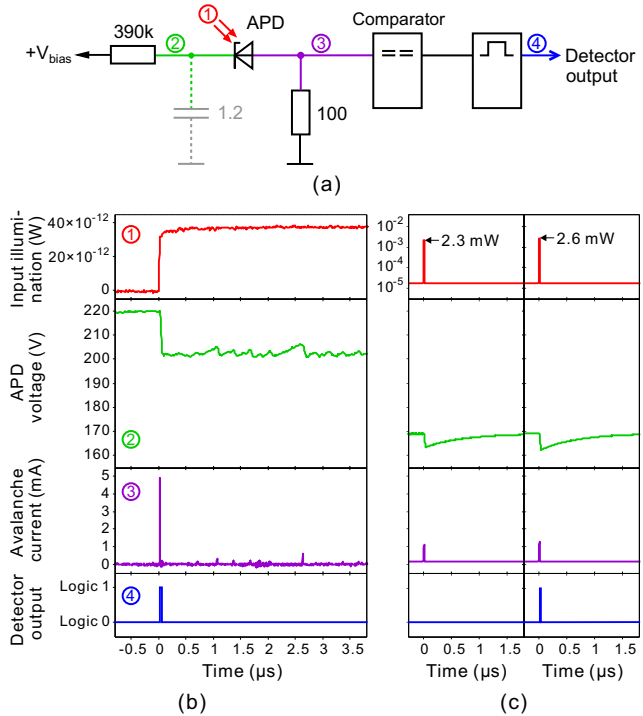}
  \caption{{\bf Detector blinding and control}. ({\bf a}) Circuit diagram of the custom-built single-photon detectors used in the QKD system under attack~\cite{marcikic:06,ling:08,peloso:09}. An avalanche photodiode (APD, PerkinElmer C30902S) is biased  $15$~V above its breakdown voltage from a voltage supply ${\rm +V_{bias}} \approx 220$~V. The avalanche current is fed by a charge stored in a small stray capacitance ($\approx1.2$~pF) and is detected via a voltage spike at the $100$~$\Omega$ resistor. The avalanche quickly self-quenches due to discharge of the capacitance and concomitant bias voltage drop; its recharge and recovery of single-photon sensitivity takes $\sim 1\,\micro\second$. ({\bf b}) Oscillograms show one of the detectors blinded after switching on $38$~pW continuous-wave (c.w$.$) illumination. ({\bf c}) Oscillograms show the same detector blinded with $17\,\micro\watt$ c.w$.$ illumination. A superimposed optical trigger pulse with a peak power of $2.3$~mW never causes a click, whereas one with ${\rm P_{th}} = 2.6$~mW always does.}
  \label{fig:singlechannel}
\end{figure}

Eve's full control of Bob's detection outcomes is crucial to the success of the faked-state attack. Several technological vulnerabilities allow for the needed degree of control~\cite{makarov:05,makarov:06,makarov:10,wiechers:10}. We have chosen to exploit blindability and controllability of single-photon detectors under strong illumination~\cite{makarov:10,lydersen:10}. The QKD system under attack uses passively quenched single-photon avalanche photodiodes (APDs, Fig.~\ref{fig:singlechannel}a). Ordinarily, the arrival of a single photon generates an electron-hole pair that leads to an avalanche in the APD. The resulting current spike is detected by a comparator and a pulse-shaper as the arrival of a single photon, a `click'. Spurious capacitances of the device result in a finite recharging time and cause a detector deadtime of $\sim 1\,\micro\second$. If the illumination level is increased such that no full recharge occurs between individual photons, the avalanche becomes progressively smaller. Under higher illumination conditions, it falls below the comparator threshold and can not be identified as a click; the detector becomes blind (Fig.~\ref{fig:singlechannel}b). Hence, by injecting high light levels into the channel, it is straightforward for Eve to indefinitely blind Bob's detectors. Under these illumination conditions, the APD no longer behaves as a single-photon detector, but as a classical photodiode generating photocurrent proportional to the optical power. A strong light pulse with peak power above a threshold ${\rm P_{th}}$ generates a current spike that mimics the signal of a legitimate photon (Fig.~\ref{fig:singlechannel}c)~\cite{lydersen:10}.

\subsection{Experimental implementation}

This QKD implementation has four detectors and utilizes a four-state protocol with polarization coding and passive basis choice (Fig.~\ref{fig:layout}b). Eve can blind all detectors using a laser diode (LD) emitting continuous-wave (c.w$.$) circularly polarized light, which splits evenly between Bob's detectors. To selectively make one detector click while keeping the other three blinded, Eve adds a linearly polarized pulse of the same polarization as the target detector, and peak power ${\rm 2P_{th}}$. By using four LDs aligned to vertical, horizontal and $\pm 45\,^{\circ}$ polarizations, Eve has the option to deliberately launch a click in any of Bob's detectors. She then executes the faked-state attack.

Before attack, we inserted Eve into the line and manually aligned her polarizations to match Bob's detector settings. Then we characterized fidelity of her control over Bob. During a $5$~min session Eve received 8,736,719 clicks and resent an equal number of faked states to Bob. Of the latter, 99.75\% caused clicks in Bob, and more importantly those clicks were always produced in the intended detector (Table~\ref{table:matrix}). Since the synchronization protocol involves Bob sending to Alice precise timing of every click registered~\cite{kurtsiefer:08}, Eve can easily identify and discard the few faked states that did not register at Bob, and that will be discarded in the reconciliation between Alice and Bob. After this, she has an identical record with Bob. Owing to small imperfections in tuning Eve's FSG (Methods Section~\ref{sec:suppl-complete-eve}), Bob had a probability of $5\times10^{-7}$ to register simultaneous clicks in two detectors, corresponding to 4 events in 323~seconds. In this QKD implementation, such double clicks were treated as noise and discarded (which is obviously insecure but easily patchable by assigning instead random bit values~\cite{lutkenhaus:99a}). We remark that our control scheme could be extended to reproduce arbitrary clicks in several detectors with a more complex FSG, which is however not needed in the present experiment.

\begin{table}\footnotesize
\caption{{\bf Fidelity of Eve's control over Bob.}} 
\label{table:matrix}\centering %
\newcolumntype{Y}{@{\extracolsep{\fill}}>{\columncolor[gray]{0.92}}c@{}}
\newcolumntype{W}{@{\extracolsep{\fill}}>{\columncolor[gray]{0.92}}c@{}}
\newcolumntype{F}{@{\extracolsep{\fill}}>{\columncolor[gray]{0.92}}p{8.6cm}}
\newcolumntype{G}{@{\extracolsep{\fill}}>{\columncolor[gray]{0.92}}p{0.15cm}}
\newcolumntype{Z}{>{\centering\arraybackslash}X}
\renewcommand{\tabularxcolumn}[1]{>{\arraybackslash}m{#1}}
\begin{tabularx}{8.6cm}{@{}p{1.4cm}@{}c@{}c@{}G@{}Z@{}Z@{}Z@{}Z@{}}
\multicolumn{3}{Y@{}}{{\bf Faked states}}                                &&\multicolumn{4}{W}{{\bf Clicks at Bob}} \\
\rowcolor[gray]{0.92}
\multicolumn{3}{Y@{}}{{\bf sent by Eve}}                                  &&V                                   &$-45\,^{\circ}$                     &H                                   & $+45\,^{\circ}$ \\
\hline
\colrule
\rowcolor[rgb]{0.85,1,0.85}
1,702,067\cellcolor[gray]{0.92}  &\multicolumn{2}{Y@{}}{{V}}               &\cellcolor[gray]{0.92}&1,693,799 99.51\%\cellcolor[gray]{1}&0                                   &0                                   &0\\
\rowcolor[rgb]{0.85,1,0.85}
2,055,059 \cellcolor[gray]{0.92} &\multicolumn{2}{Y@{}}{{$-45\,^{\circ}$}} &\cellcolor[gray]{0.92}&0                                   &2,048,072 99.66\%\cellcolor[gray]{1}&0                                   &0\\
\rowcolor[rgb]{0.85,1,0.85}                
2,620,099\cellcolor[gray]{0.92}  &\multicolumn{2}{Y@{}}{{H}}               &\cellcolor[gray]{0.92}&0                                   &0                                   &2,614,918 99.80\%\cellcolor[gray]{1}&0\\
\rowcolor[rgb]{0.85,1,0.85}
2,359,494\cellcolor[gray]{0.92}  &\multicolumn{2}{Y@{}}{{$+45\,^{\circ}$}} &\cellcolor[gray]{0.92}&0                                   &0                                   &0                 &2,358,418 99.95\%\cellcolor[gray]{1}\\
\hline
\colrule
\rowcolor[gray]{0.92}
&&&&&&&\\
\multicolumn{8}{F@{}}{The $4\times4$ matrix shows the total number of clicks in each of Bob's detectors as well as their percentage in respect to the faked states sent with the same polarization. The data was recorded during a $5\,\minute$ diagnostic-mode session. The lack of off-diagonal elements proves that a click is never launched in a wrong detector. Double clicks are not included. The overall click rate is close to 100\%, leading to virtually no loss in the line Eve--Bob.}\\
\end{tabularx}
\end{table}

\subsection{QKD performance and key extraction}

\begin{figure}
  \includegraphics[width=8cm]{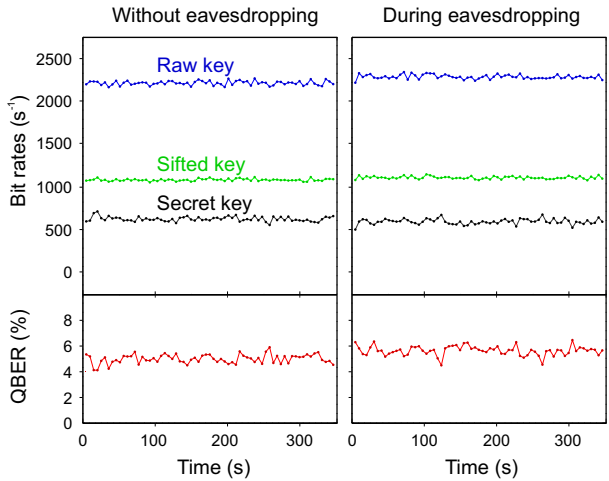}
  \caption{{\bf QKD performance with and without eavesdropping as measured by Alice and Bob}. Session without Eve in the fibre line (left). Eve installed (right). The traces in the top chart correspond to the raw key rate, sifted key rate and final secret key rate after error correction and privacy amplification~\cite{marcikic:06,peloso:09}. The bottom chart shows the quantum bit error ratio (QBER).}
  \label{fig:result}
\end{figure}

After Eve's calibration, we ran multiple 5--10~min QKD sessions over a few hours, some with Eve inserted in the fibre line and some without. We recorded performance statistics, all public communication data between Alice and Bob, and the generated keys. During QKD, the legitimate parties monitor key rates to check the line transmission. Fig.~\ref{fig:result} shows results from two typical sessions, one eavesdropped and one not. As expected, inserting Eve does not alter the rates. Small differences in rate averages of the two sessions are not caused by eavesdropping but rather are normal medium-term alignment fluctuations in this QKD system. The quantum bit error ratio (QBER) of 5--6\% is typical for this experiment~\cite{marcikic:06,ling:08,peloso:09}, and well below the security limit for the Bennett-Brassard-Mermin 1992 (BBM92) protocol used here~\cite{scarani:09}.

In the sessions where Eve was connected, she extracted Bob's sifted key from her clicks and the recorded public communication Alice--Bob. Alice and Bob identify photon pairs by time-tagging each detector click and exchanging these times over the public channel~\cite{kurtsiefer:08}. This allows them to synchronize their clocks and to keep track of what photons were detected. Bob also announces his detection bases, and Alice answers for which Bob's clicks she detected the other photon of the pair in the same basis (these pairs form the sifted key). Since no measurement outcomes are revealed, this information can be entirely public. In the present implementation, this channel is established over a transmission control protocol and internet protocol (TCP/IP) wireless connection, and is passively wiretapped by Eve. She watches the discussion, synchronizes her clock with Bob's clock, then sifts her key keeping only those of her clicks which are also kept by Alice and Bob in the sifted key. We ran Eve's processing script on recorded experimental data and verified that {\it in all eavesdropped QKD sessions, Eve's sifted key was identical to Bob's} (the script and data sample are available, Methods Section~\ref{sec:suppl-raw-data}).

If the source, analysers and transmission medium were perfect, this sifted key would directly constitute the secret key. Under realistic conditions, the sifted keys of Alice and Bob are not identical (the difference being quantified by the quantum bit error ratio). Further steps of error correction and privacy amplification complete the public exchange Alice--Bob to produce the secret key~\cite{bennett:92a,scarani:09}. Since Eve has the same sifted key as Bob, she can apply the same processing as Bob to it, and is guaranteed to produce the same secret key.

\section*{Discussion}

The particular weakness exploited in this work can be closed by developing suitable countermeasures~\cite{lydersen:11}. The incoming blinding light may be detected, either by a separate watchdog detector or by monitoring electrical and thermal parameters of the APDs. Single-photon sensitivity of Bob's APDs can be tested at random times by a calibrated light source placed inside Bob. The eavesdropper introduces 212~ns time delay (Methods Section~\ref{sec:suppl-delay}), however monitoring may be impractical, and Eve can compensate this delay by shortening the fibre line. Eve's need to calibrate her FSG before the attack cannot be considered a reliable deterrent, because she may calibrate non-obtrusively~\cite{makarov:05}. Other countermeasure proposals that break the described attack exist and may be relatively easy to implement. However a countermeasure that incorporates into the existing security proofs~\cite{scarani:09,gottesman:04,fung:09,maroy:10}, such as the one in ref.~\onlinecite{lydersen:11}, has not yet been implemented.

In conclusion, we have demonstrated a complete and undetected eavesdropping attack against an established QKD system. The success of this demonstration proves that a technological imperfection in a QKD system can be fully exploited using off-the-shelf components. As there is a variety of potentially exploitable loopholes in both research and commercial QKD systems~\cite{vakhitov:01,kurtsiefer:01a,makarov:05,makarov:06,lamas-linares:07a,zhao:08,nauerth:09,makarov:10,xu:10,lydersen:10,wiechers:10}, Eve can design a tailored attack on one or the other implementation problem. We have briefly discussed how one particular loophole can be closed. However, a more pointed question is what problems still lurk unnoticed in the gap between the theoretical description and the practical systems~\cite{scarani2009a}. Just as in classical cryptography, an ongoing search for backdoors is required to build hardened implementations of quantum cryptography for real-world use.

\section*{Methods}

\setcounter{subsection}{0}

\subsection{\label{sec:suppl-complete-eve}Complete Eve's setup}

\begin{figure}[t]
  \includegraphics[width=8.6cm]{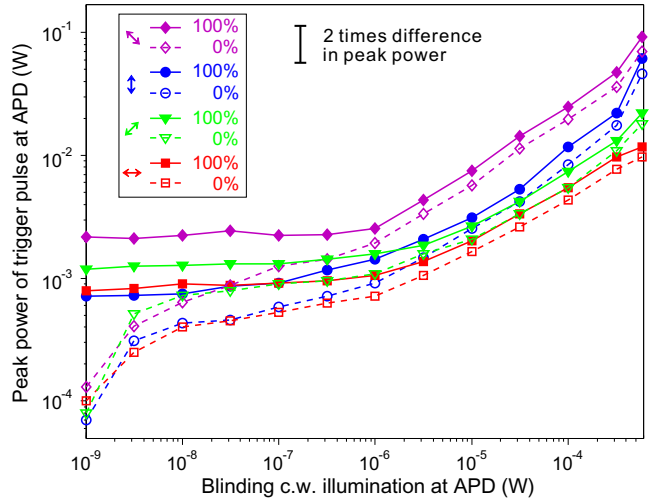}
  \caption{{\bf Click thresholds of the four detectors in Bob's PA versus blinding c.w$.$ power.} The dashed curves show the highest peak pulse power at which the detector still never clicks. The solid curves show the lowest peak pulse power at which it always clicks. Between these two thresholds, click probability of each detector increases gradually. The detector recording photons of horizontal polarization (curves with squares) was the one tested in Fig.~\ref{fig:singlechannel}.}
  \label{fig:detectors_response}
\end{figure}

The task of Eve's FSG is to make the target detector at Bob click, while keeping his other detectors silent. An optical pulse of a peak power ${\rm P_{th}}$ at the target detector causes it click with 100\% probability. In order for the FSG depicted in Fig.~\ref{fig:layout}b to work, a pulse of power ${\rm P_{th}}/2$ should never cause the two conjugate-basis detectors to click. Unfortunately, for the actual Bob's polarization analyser (PA) this condition did not hold, because one of its detectors turned out to have significantly higher click thresholds than the other three (see Fig.~\ref{fig:detectors_response}). Note that for blinding power $>$$1\,\micro\watt$, the click thresholds of all four detectors rose uniformly. We tried to change the circular blinding polarization to elliptical, such that the detector with higher click threshold received much less blinding power than the other three. This achieved {\it almost} perfect fidelity of Eve's control over Bob, with diagonal elements $>$96.2\% (in terms of Table~\ref{table:matrix}) and off-diagonal elements $<$0.005\%. The latter meant Eve had slightly less than full information on the sifted key, compromising the security but requiring an additional cryptanalytic task to complete the eavesdropping.

\begin{figure*}[htpb]
  \includegraphics[width=16.5cm]{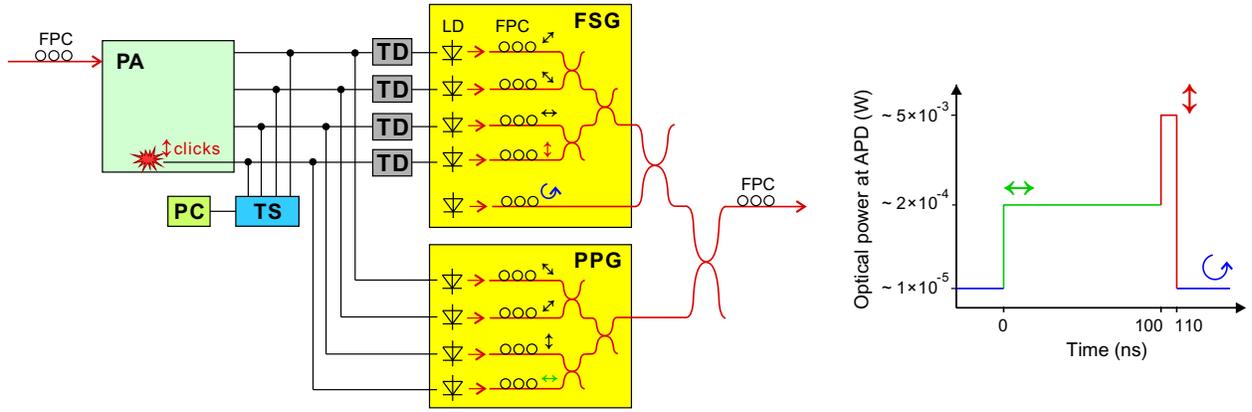}
  \caption{{\bf Complete Eve's setup forming an improved control diagram.} This setup was used for all eavesdropping experiments reported in this article. Four LDs followed by FPCs comprise a pre-pulse generator (PPG). Timing of each main trigger pulse is adjusted by a trimmable time-delay (TD) circuit. The FPC before the PA is used to align Eve's polarization reference frame after inserting her into the line (similarly to the manual FPC used in Alice's setup for alignment during QKD system installation).}
  \label{fig:eve_actual}
\end{figure*}

We then improved the control method by including a polarized pre-pulse that dynamically increased blinding power at the orthogonal-basis detectors $100\,\nano\second$ before the main trigger pulse was sent (Fig.~\ref{fig:eve_actual}). These pre-pulses were emitted by four additional laser diodes. With this setup, clicks never occurred in a wrong detector. When we calibrated Eve's control of Bob by sending the same faked state at a fixed rate, the click probability in any target detector was 100\%, and double clicks did not occur. However as we discovered later in the recorded experimental data, a cross-talk between adjacent faked states (which could be as closely spaced as $550\,\nano\second$ during eavesdropping) led to slightly less than 100\% click probability, as Table~\ref{table:matrix} illustrates. There were also a few double clicks. Nevertheless Eve managed to recover complete sifted key by proper post-processing, which shows robustness of this control method.

\subsection{\label{sec:suppl-delay}Jitter and insertion delay introduced by Eve}

\begin{figure}[t]
  \subfigure[]{\includegraphics[width=8.4cm]{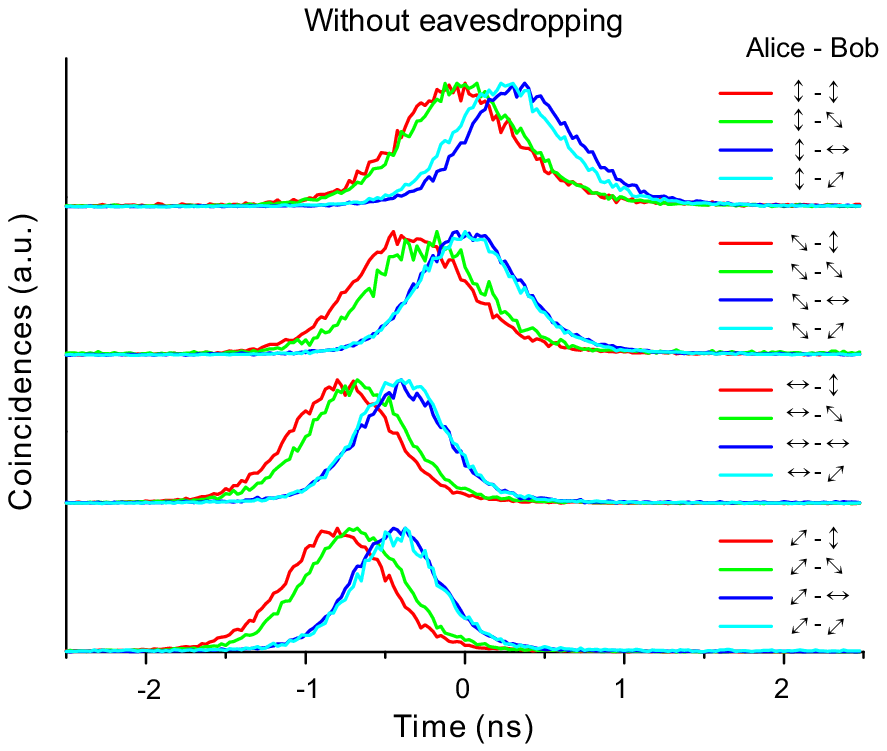}}
  \subfigure[]{\includegraphics[width=8.4cm]{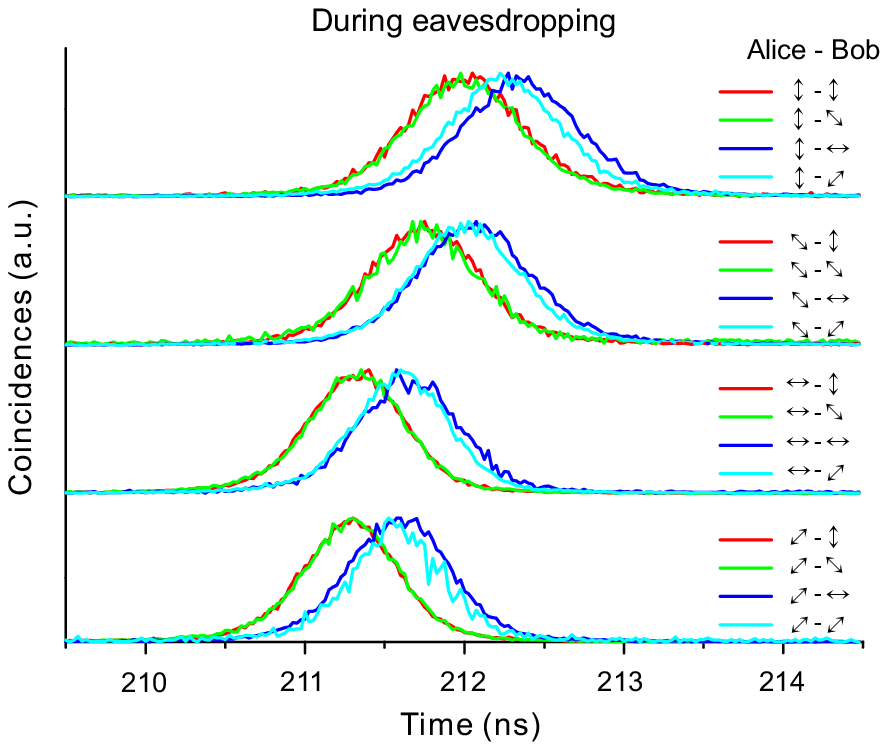}}
  \caption{{\bf Coincidence timing distributions for all 16 detector combinations Alice--Bob,} measured at outputs of Alice's and Bob's detectors and normalized. ({\bf a}) Distributions without eavesdropping. The full-width at half-magnitude (FWHM) averaged over the 16 coincidence peaks is $761\,\pico\second$. The QKD software corrects for relative time shift between the peaks via a set of calibration constants. ({\bf b}) Distributions with Eve inserted into the line, after TD adjustment. The FWHM averaged over the 16 peaks is $779\,\pico\second$.}
  \label{fig:histo}
\end{figure}

After initially inserting Eve into the line, her four detection and Bob control channels had slightly different insertion delays (varying by {\small $\lesssim$}$1\,\nano\second$). Since Alice and Bob used a tight coincidence window to identify photon pairs, we had to equalize Eve's insertion delays by adjusting the time-delay circuits (shown in Fig.~\ref{fig:eve_actual}). As can be seen in Fig.~\ref{fig:histo}, the resulting relative coincidence time distributions were indistinguishable from those without eavesdropping. The jitter between photon pairs stayed about the same and was dominated by timing jitter of the single-photon detectors, $\approx$$500\,\pico\second$ full-width at half-magnitude for each detector.

As Fig.~\ref{fig:histo} shows, Eve introduced an overall insertion delay of $212\,\nano\second$. This went without any consequence, because Alice and Bob synchronized their clocks by photon coincidences, which is a common practice in QKD systems of this type. In general, the propagation delay is not authenticated and is not a part of the QKD security. We remark that if Alice and Bob synchronized their clocks in some independent way (which is probably impractical), Eve could cancel her insertion delay by shortening the fibre-optic line and/or bypassing a part of the line by spatially separating her polarization analyser and FSG and establishing a line-of-sight radio-frequency link between them, in which signals travel $\sim$1.5 times faster than in fibre~\cite{makarov:05}. These tricks would not apply to systems using a free-space line-of-sight QKD link~\cite{rarity:01,hughes:02,kurtsiefer:02,ursin:07,marcikic:06,ling:08,peloso:09}, but so far none of them implemented a clock synchronization method that would fail because of Eve's insertion delay.

\subsection{\label{sec:suppl-raw-data}Raw experimental data and Eve's key extraction software}

There were four eavesdropped QKD sessions over $2\,\hour$. For example, the second session lasted $5\,\minute$ and produced 393,323-bit sifted key, identical between Bob and Eve. The raw data recorded during this session and the script used to extract Eve's sifted key can be found in a single archive file: {\it \nohyphens{http://www.vad1.com/eve-extract-sifted-key.zip}} (74~MiB). The minimum disk space required is 125~MiB, including files generated by running the script.

The main script to do Eve's key extraction, named {\it \nohyphens{eve$\_$extract$\_$sifted$\_$key.m}}, can be found in the directory {\it \nohyphens{scripts-matlab}}, while the other files in this directory are functions called by the main script, and a log file {\it \nohyphens{proclog.txt}} will be generated after running the script. The script is written in MATLAB. We have tested it under both Windows and Linux.

The directory {\it \nohyphens{data-raw}} contains the raw experimental data from this session, recorded during the experiment. To obey realistic eavesdropping conditions, Eve only gets access to the classical channel where the transmission is public (and to her own computer), but not to Bob's or Alice's computers. Hence, the script is run only upon the timing and basis choice data sent from Bob to Alice (the subdirectory {\it \nohyphens{alice-receivefiles}}), the sifting response returned from Alice (the subdirectory {\it \nohyphens{bob-receivefiles}}), and Eve's own recorded click data (the subdirectory {\it \nohyphens{eve-raw-events}}). Although not used by the extraction script, both sifted and final secret keys recorded in Alice's and Bob's computers are also provided in the archive, to satisfy a curious reader. The final secret key is 218,462 bit long.

After running the script, Eve's sifted key will be extracted and stored in a new directory named {\it \nohyphens{data-produced-by-scripts}}. The script then does a bitwise comparison between Eve's and Bob's sifted keys, and reports the number of discrepancies (which is zero for all eavesdropped QKD sessions). For convenience, both Bob's and Eve's sifted keys are also saved as two sets of ASCII files.

All data is partitioned into files by {\it epoch} (defined as a time span of $2^{29}\,\nano\second \approx 0.537\,\second$), except the final secret key which is stored in blocks of 9 epochs. All file formats are openly defined and documented~\cite{kurtsiefer:08}, and have been used in several QKD experiments previously~\cite{marcikic:06,ling:08,peloso:09}.

\section*{Acknowledgements}
This work was supported by the National Research Foundation and the Ministry of Education, Singapore, and the Research Council of Norway (grant no$.$ 180439/V30). L$.$ Lydersen and V$.$ Scarani are thanked for useful discussions. We thank the OLPC project for providing a notebook for the eavesdropper.

\section*{Author contributions}
V{.}M{.} conceived the idea. Q{.}L{.}, I{.}G{.}, A{.}L{.}-L{.}, C{.}K. and V{.}M{.} prepared and conducted the experiment. Q{.}L{.} and A{.}L{.}-L{.} processed the recorded data with help of I{.}G{.} and C{.}K.~ Q{.}L{.}, A{.}L{.}-L{.}, I{.}G{.} and V{.}M{.} wrote the paper. J{.}S{.} supervised the NTNU team. C{.}K{.} and V{.}M{.} supervised the project.

\section*{Additional information}
The authors declare no competing financial interests.

\end{document}